\documentclass[aps,%twocolumn
showpacs,superscriptaddress,citeautoscript,reprint,pra,floatfix
]{revtex4-1}

\usepackage{bm}
\usepackage[dvipdfmx]{graphicx}
\usepackage{bmpsize}
\usepackage{amsmath}
\usepackage{amsfonts}
\usepackage{units}
\usepackage{color}
\usepackage[colorlinks=true,%
            linkcolor=blue,%
            urlcolor=blue,%
            citecolor=blue,%
            filecolor=blue,%
            bookmarksopen=true,%
            pdfauthor={DZ_JPRA_PAO},%
            pdftitle={MBS_BIC_MQDS},%
            pdfsubject={MBS_BIC_MQDS_Manuscript},%
            pdfpagemode=UseOutlines]{hyperref}

\definecolor{Blue}{rgb}{0.3,0.3,0.9}
\definecolor{Red}{rgb}{0.9,0.3,0.3}
\definecolor{Green}{rgb}{0.3,0.6,0.3}
\definecolor{Black}{rgb}{0.0,0.0,0.0}

\begin{document}

\title[Tunable single-photon routers]%
{%Waveguide resonators coupled to cavities as a tunable single photon quantum router\\
A tunable single photon quantum router}

\author{M.~Ahumada}
%\email[]{mahumadacortes@gmail.com} 
\affiliation{Departamento de F\'{\i}sica, Universidad T\'{e}cnica 
Federico Santa Mar\'{\i}a, Casilla 110 V, Valpara\'{\i}so, Chile}

\author{P.~A.~Orellana}
%\email[]{pedro.orellana@usm.cl} 
\affiliation{Departamento de F\'{\i}sica, Universidad T\'{e}cnica 
Federico Santa Mar\'{\i}a, Casilla 110 V, Valpara\'{\i}so, Chile}

\author{F.~Dom\'{\i}nguez-Adame}
%\email[]{adame@ucm.es}
\affiliation{GISC, Departamento de F\'{\i}sica de Materiales, Universidad Complutense, E-28040 Madrid, Spain}

\author{A.~V.~Malyshev}
%\email[]{a.malyshev@fis.ucm.es}
\affiliation{GISC, Departamento de F\'{\i}sica de Materiales, Universidad Complutense, E-28040 Madrid, Spain}
\affiliation{Ioffe Physical-Technical Institute, 26 Politechnicheskaya str., 194021 St.-Petersburg, Russia}


\begin{abstract}

We propose an efficient single-photon router comprising two resonator waveguide channels coupled by several sequential cavities with embedded three-level atoms. We show that the system can operate as a perfect four-way single-photon switch. We also demonstrate that an incident single-photon propagating in one of the waveguides can be routed into one or the other output channels; such routing can be controlled by the external classical electromagnetic field driving the atoms. We argue that, under appropriate conditions, the efficiency of such routing can be close to 100\% within a broad operational bandwidth, suggesting various applications in photonics.

\end{abstract}

\maketitle

\section{Introduction} \label{sec:intro}

In the last years, there has been significant progress in the control of hybrid light-matter systems at the frontier between quantum optics and physics of condensed matter. Some examples of these hybrid quantum systems are electrodynamic cavities, cold atoms coupled to light, opto-mechanical devices, and atoms embedded in quantum cavities~\cite{QEDRev,RevModPhys.85.623,RevModPhys.87.1379,RevModPhys.89.021001}. The coupled resonator waveguides (CRWs) provide a platform to study the light-matter interaction with high precision~\cite{Notomi2008}. Atoms in CRWs circuits bring the possibility to investigate the photonic quantum transport with very high sensitivity.  Photons, in comparison with other possible information carriers such as electrons, can sustain quantum coherence for vast distances, which makes them excellent candidates for transferring and manipulating quantum information~\cite{Monroe2002,Northup2014,vanLoo1494,Ritter2012}. Hence single-photon transport through CRWs has received considerable attention in the last decade. 

One of the most relevant devices for the operation of a quantum network is a quantum router (QR), whose primary function in the simplest configuration is to send or route an incident photon into one of the two output channels~\cite{Kimble2008}. Recently, there have been several theoretical and experimental proposals for quantum routers based on several different structures, such as CRWs~\cite{PhysRevLett.111.103604,PhysRevA.89.013805,Huang2018,J.Huang,Lu:15,Liu2016}, whispering gallery resonators~\cite{Aoki2009,PhysRevX.3.031013,Shomroni903,Li2016,Cao:17}, waveguide-emitter system~\cite{Yan2014,PhysRevA.97.023821}, superconducting qubit~\cite{PhysRevLett.107.073601} and quantum electrodynamics system~\cite{Yuan2015,Hu2017}. 
In the latter context, Zhou \emph{et al.}~\cite{PhysRevLett.111.103604,PhysRevA.89.013805} proposed an experimentally accessible single-photon routing scheme comprising two quantum channels connected by a resonant cavity with a single-type three-level atom embedded into it. It was demonstrated that the output channel for a propagating wave packet could be selected by applying a classical electromagnetic field to the atom. Based on the above mentioned works, several proposals have emerged, such as quantum memories and quantum gates~\cite{Huang2018,Li2016} to name a few examples. However, all these proposals have considerable limitations such as relatively low efficiency of switching between output channels and narrow operational bandwidth. 

In the present paper, we propose a device design that circumvents the limitations mentioned above. We consider two CRWs coupled by several sequential cavities with embedded three-level atoms. We demonstrate that such a device can operate in two different modes: (i)~An incident photon is routed into one of the four channels with equal transmission probability of $1/4$ and (ii)~one of the two output channels is selected by the external classical electromagnetic field driving the atoms. In the latter case, the transmission probability in the selected channel is close to unity within a broad band of photon energies and a wide range of parameters.

\section{Model} \label{sec:Model}

\begin{figure}[ht]
\includegraphics[width=\columnwidth]{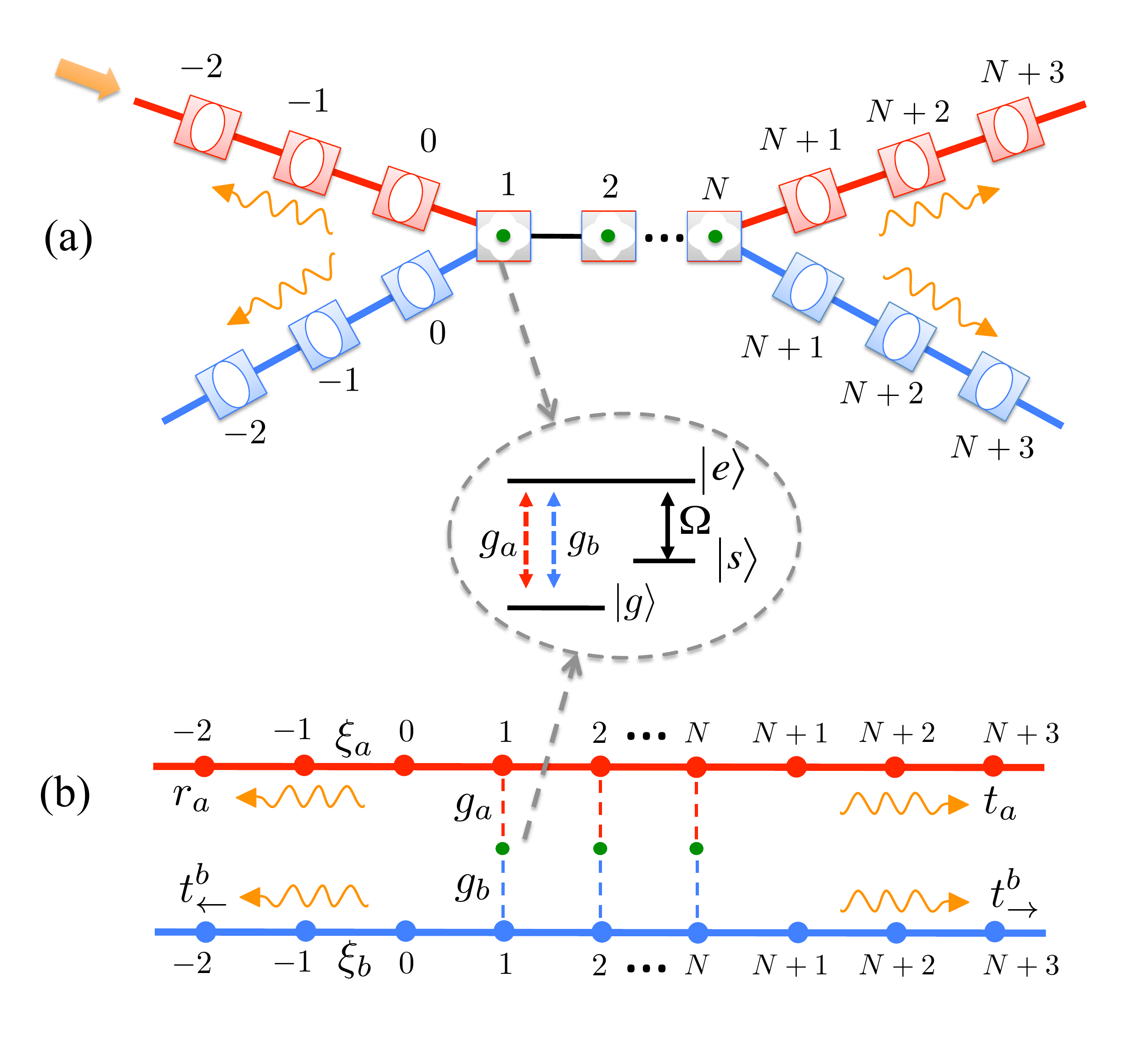}
\caption{Schematic view of the single-photon router. (a) Two CRWs with nearest-neighbor inter-coupling. The CRWs are intra-coupled by $N$ cavities with embedded 
three-level atoms. The $N$ interconnected cavities constitute the scattering region of the device. The single photon (represented by the yellow arrow) impinges on the scattering region from the left arm of channel $a$. The inset shows the atomic level scheme. The transition between the ground and excited state $\vert g_{j}\rangle\leftrightarrow\vert e_{j}\rangle$ are dipole-coupled to the $j$th cavity mode of CRW-$a$ and CRW-$b$  with strength $g_{a}$ and $g_{b}$, respectively. The transition between the excited and the third state $\vert e_{j}\rangle\leftrightarrow\vert s_{j}\rangle$ is driven by an external controlling field with Rabi frequency $\Omega$. (b) Simplified scheme of the router. The amplitudes of the transmission $t_a$, reflection $r_a$, transfer $t^{b}_{\leftarrow}$ and, $t^{b}_{\rightarrow}$ are represented by yellow wavy arrows.
}
\label{Fig1}
\end{figure}

Our proposed system is depicted schematically in Fig.~\ref{Fig1}. The system comprises two channels: CRW-$a$ and CRW-$b$ (shown as red and blue chains in Fig.~\ref{Fig1}), each being a quasi-one-dimensional array of identical optical cavities with nearest-neighbor coupling. A section of $N$ sequential sites of the two waveguides (numbered $1,2,\ldots, N$) are coupled via cavities with embedded three-level atoms. Each atom has a ground state $\vert g\rangle$, an excited state $\vert e\rangle$ and a third state $\vert s\rangle$. The transition $\vert g\rangle\leftrightarrow\vert e\rangle$ of each of the atoms is dipole-coupled to the cavity modes of the nearest CRW--$a$ and CRW--$b$ with coupling strengths $g_{a}$ and $g_{b}$, respectively. The atomic transition $\vert g\rangle\leftrightarrow\vert s\rangle$ is forbidden. Finally, an external classical controlling field of frequency $\nu$ drives the transition $\vert g\rangle\leftrightarrow\vert e\rangle$ with the Rabi frequency $\Omega$. 

The total Hamiltonian of the system can be split into three terms, $H=H_\mathrm{ab}+H_\mathrm{A}+H_\mathrm{int}$. $H_\mathrm{ab}$ describes the photon propagation through the CRW--$a$ and CRW--$b$ and is given by the tight-binding bosonic model. $H_\mathrm{A}$ is the free Hamiltonian of the three-level atoms. $H_\mathrm{int}$ describes the atom, field cavities, and classical field interaction. A Jaynes-Cummings Hamiltonian represents this interaction term under the rotating wave approximation. Thus, the various terms of the Hamiltonian are given by ($\hbar=1$ in what follows)
\begin{eqnarray}
H_\mathrm{ab}&=&\sum_{i=-\infty}^{\infty}\left[\omega_{a}\hat{a}^{\dag}_{i}\hat{a}_{i}-\xi_{a}
(\hat{a}^{\dag}_{i}\hat{a}_{i+1}+\hat{a}^{\dag}_{i+1}\hat{a}_{i})\right]\,\nonumber\\
&+&\sum_{i=-\infty}^{\infty} \left[\omega_{b}\hat{b}^{\dag}_{i}\hat{b}_{i}-\xi_{b} (\hat{b}^{\dag}_{i}\hat{b}_{i+1}+\hat{b}^{\dag}_{i+1}\hat{b}_{i})\right]\ ,\nonumber \\
H_\mathrm{A}&=&\sum_{j=1}^{N}\Big[\omega_{e}\vert e_{j}\rangle\langle e_{j}\vert+\omega_{s}\vert s_{j}\rangle\langle s_{j}\vert\Big]\ ,\nonumber\\
\nonumber 
H_\mathrm{int}&=&\sum_{j=1}^{N}\Big[\vert e_{j}\rangle\langle g_{j}\vert(g_{a}\hat{a}_{j}+g_{b}\hat{b}_{j})+\Omega\vert e_{j}\rangle\langle s_{j}\vert e^{-i\nu t} \nonumber\\
&+& \mathrm{H.c.}\Big]\ ,
\end{eqnarray}
where $\hat{a}^{\dag}_{i}$ ($\hat{a}_{i}$) and $\hat{b}^{\dag}_{i}$ ($\hat{b}_{i}$) are the creation (annihilation) operators of a single photon in the $i$th cavity of CRW--$a$ and CRW--$b$ with frequencies $\omega_{a}$ and $\omega_{b}$, respectively.
$\omega_{s}$ and $\omega_{e}$ are the third and excited state frequencies, respectively.
$\xi_{a}$ and $\xi_{b}$ are the nearest-neighbor couplings for the waveguide $a$ and $b$. Here $\mathrm{H.c.}$ stands for Hermitian conjugate. The dispersion relation for the CRW--$a$ and CRW--$b$ are given by $E_{a}=\omega_{a}-2\,\xi_{a}\,\cos k_{a}$ and $E_{b}=\omega_{b}-2\,\xi_{b}\,\cos k_{b}$, resulting in energy bands with bandwidth $4\xi_{a}$ and $4\xi_{b}$, respectively. 

We consider the single-photon scattering process in the rotating frame. To this end, we perform a unitary transformation
\begin{eqnarray}
H^{\prime}&=&U^{\dag} H U-i\,U^{\dag}\,\dfrac{\partial}{\partial t }\,U\ ,\nonumber \\
U&=&\prod_{j=1}^{N} e^{i\nu t \vert s_{j}\rangle\langle s_{j}\vert}\ ,
\end{eqnarray}
which turns $H$ into a time-independent Hamiltonian $H^{\prime}=H_\mathrm{ab}+H^{\prime}_{A}+H^{\prime}_{int}$ with
\begin{eqnarray}
H^{\prime}_\mathrm{int}&=&\sum_{j=1}^{N}\Big[\vert e_{j}\rangle\langle g_{j}\vert(g_{a}\hat{a}_{j}+g_{b}\hat{b}_{j})
+\Omega\vert e_{j}\rangle\langle s_{j}\vert\Big] + \mathrm{H.c.}\ ,
\nonumber\\
H^{\prime}_{A}&=&\sum_{j=1}^{N}\Big[\omega_{e}\vert e_{j}\rangle\langle e_{j}\vert+\omega^{\prime}_{s}\vert s_{j}\rangle\langle s_{j}\vert\Big]\ ,
\end{eqnarray}
where $\omega^{\prime}_{s}=\omega_{s}+\nu$. $H_\mathrm{ab}$ remains invariant under this transformation. 

\section{Single photon scattering}

The propagation of a single photon through the system can be assessed by inspecting the energy spectrum of the Hamiltonian $H^{\prime}$. This can be obtained by expressing the single excitation eigenstate as
\begin{eqnarray}
\vert \psi_{E}\rangle&=&\sum_{i=-\infty}^{\infty}\Big[\alpha(i)\hat{a}^{\dag}_{i}\vert 0,g\rangle+\beta(i)\hat{b}^{\dag}_{i}\vert 0,g\rangle \\ \nonumber
&+&\sum_{j=1}^{N} u_{e,j}\vert 0,e_{j}\rangle + u_{s,j}\vert 0,s_{j}\rangle\Big]\ .
\end{eqnarray}
Here $\alpha(i)$ and $\beta(i)$ are the probability amplitudes to find the photon in the $i$th cavity of CRW--$a$ and CRW--$b$, respectively. $u_{e,j}$ and $u_{s,j}$ are the probability amplitudes of the $j$th three-level system in the excited and third state, respectively, and $\vert 0\rangle$ is the vacuum state of the CRWs.

We obtain the following coupled stationary equations for the amplitudes from the eigenvalue equation $H\vert \psi_{E}\rangle=E\vert \psi_{E}\rangle$
\begin{subequations}
\begin{eqnarray} 
&&\hspace{-5mm}(E-\omega_{e})u_{e,j}=\Omega u_{s,j}+g_{a,j}\,\alpha(j)+g_{b,j}\,\beta(j)\ ,\nonumber\\  
&&\hspace{-5mm}(E-\omega_{s}) u_{s,j}=\Omega^{\ast} u_{e,j}\ ,\nonumber\\
&&\hspace{-5mm}(E-\omega_{a})\,\alpha(j)=\xi_{a}[\alpha(j+1)+\alpha(j-1)]+g_{a,j}\,u_{e,j}\ ,\nonumber\\
&&\hspace{-5mm}(E-\omega_{b})\,\beta(j)=\xi_{b}[\beta\,(j+1)+\beta(j-1)]+g_{b,j}\,u_{e,j}\ ,
\label{linear Eqs.}
\end{eqnarray}
where
\begin{equation}
g_{a (b),j}=\ \begin{cases}
   0\ ,\qquad j<1,\ j>N\ ,\\
  g_{a (b)}\ ,\qquad 1\leq j \leq N\ .
\end{cases}
\end{equation}
\end{subequations}
From~(\ref{linear Eqs.}) we obtain the following coupled equations
\begin{eqnarray}
(E-\widetilde{\omega}_{a,j})\alpha(j)=-\xi_{a}\left[\alpha(j+1)+\alpha(j-1)\right]+G_j(E)\,\beta(j)
\nonumber \\
(E-\widetilde{\omega}_{b,j})\beta(j)=-\xi_{b}\left[\beta(j+1)+\beta(j-1)\right]+G_j(E)\,\alpha(j)\nonumber \\
\label{Eq coupled1}
\end{eqnarray}
where $\widetilde{\omega}_{a (b),j}(E)=\omega_{a(b)}+g_{a (b),j}^{2}\,V(E)$ and 
\begin{align}
\nonumber
& G_j(E)=g_{a,j}\,g_{b,j} V(E)\ ,\\
& V(E)=\frac{E-\omega_{s}}{(E-\omega_{s})(E-\omega_{e})-\vert\Omega\vert^{2}}\ .
\label{V}
\end{align}

Under the standard scattering boundary conditions: a plane wave incident from $-\infty$ in the CRW--$a$ (see Fig.~\ref{Fig1}(b)), the photon amplitudes in the two channels can be written as: 
\begin{align}
\alpha(j)&=\ \begin{cases}
    e^{ik_{a}j}+r_{a}\,e^{-ik_{a}j}\ ,\qquad j<1\ ,\\
   t_{a}\,e^{ik_{a}j}\ ,\qquad j>N\ .
\end{cases}
\nonumber\\
\beta(j)&=\ \begin{cases}
   t^{b}_{\leftarrow}\,e^{-ik_{b}j}\ ,\qquad j<1\ ,\\
   t^{b}_{\rightarrow}\,e^{ik_{b}j}\ ,\qquad j>N\ ,
\end{cases} 
\label{sol. onda plana} 
\end{align}
where $r_{a}$ and $t_{a}$ are the reflection and transmission amplitudes in the channel $a$, while $t^{b}_{\leftarrow}$ and $t^{b}_{\rightarrow}$ being the backward and forward transfer amplitudes into the channel $b$, respectively [see Fig.~\ref{Fig1}(b)]. 

Hereafter we address the seemingly most favorable case of the maximum overlap between the energy bands of the two CRWs: setting ${\omega}_{a}={\omega}_{b}=\omega_{0}$ and $\xi_{a}=\xi_{b}=\xi$. The nearest-neighbor coupling $\xi$ will be used as a unit of energy throughout the paper. Additionally, we consider equal atom-to-CRW mode couplings $g_{a,j}=g_{b,j}=g$. 

In order to solve Eq.~(\ref{Eq coupled1}) the following transformation is performed: instead of considering the photon amplitudes $\alpha(j)$ and $\beta(j)$ in the physical channels $a$ and $b$, we consider the symmetric ($\mathcal{S}$) and antisymmetric ($\mathcal{A}$) linear combinations of them: $\psi^{\pm}(j)=\alpha(j)\pm\beta(j)$. In the $\mathcal{S-A}$ representation, Eqs.~(\ref{Eq coupled1}) reduce to
\begin{eqnarray}\label{Eq coupled2}
\nonumber 
&&(E-\varepsilon^{+})\psi^{+}(j)=-\xi[\psi^{+}(j+1)+\psi^{+}(j-1)]\ , \\ 
&&(E-\varepsilon^{-})\psi^{-}(j)=-\xi[\psi^{-}(j+1)+\psi^{-}(j-1)]\ ,
\end{eqnarray}
where the  effective site energies are $\varepsilon^{+}_{j}=\omega_{0}+2G_{j}(E)$ and $\varepsilon^{-}=\omega_{0}$.

In the $\mathcal{S-A}$ representation, the scattering boundary conditions are written as 
\begin{align}
\psi^{+}(j)&=\ \begin{cases}
    e^{ik_{+}j}+r_{+}\,e^{-ik_{+}j}\ ,\quad & j<1\ ,\\
   t_{+}\,e^{ik_{+}j}\ ,\quad & j>N\ .
\end{cases}
\nonumber\\
\psi^{-}(j)&=\ \begin{cases}
   e^{ik_{-}j}+r_{-}\,e^{-ik_{-}j}\ ,\quad & j<1\ ,\\
   t_{-}\,e^{ik_{-}j}\ ,\quad & j>N\ .
\end{cases} 
\label{sol. onda plana2} 
\end{align}
where $t_{\pm}$ and $r_{\pm}$ are the transmission and reflection amplitudes in the virtual $\mathcal{A-S}$ channels. From Eqs. (\ref{Eq coupled2}) evaluated at the boundary of the scattering region ($j=0$, $j=1$, $j=N$ and $j=N+1$) along with Eqs. (\ref{sol. onda plana2}), one can obtain closed expressions for the transmission $t_{\pm}$, and reflections $r_{\pm}$. As $\mathcal{A}$-channel is equivalent to a free channel with energy $\omega_0$, then the incident wave is transmitted without reflection and unity transmission amplitude, i.e. $r_{-}=0$ and $t_{-}=1$. Then, $r_{+}$ and $t_{+}$ are obtained
\begin{align}
&r_{+}=\frac{e^{ik}(\cos k-\cos k_{+})}{\cos k\cos k_{+}-1-i\,\cot(Nk_{+})\sin  k\sin k_{+}}\ ,\nonumber\\ 
&t_{+}=\frac{e^{-ikN}\sin k\sin k_{+}}{\sin k\sin k_ {+}\cos(Nk_ {+})\!+\!i(\cos k\cos k_{+}\!-\!1)\sin(Nk_{+})}\, ,\nonumber \\ 
&k_{+}=\arccos\left(-\frac{E-\varepsilon^{+}}{2\xi}\right)\ , \quad \textup{if}\quad  |\frac{E-\varepsilon^{+}}{2\xi}| \leq 1  \, .
\end{align}

Once transmission and reflection amplitudes in the virtual $\mathcal{S}$ and $\mathcal{A}$ channels are known, one can obtain these quantities for the physical channels $a$ and $b$ in the following way:
\begin{align}
&r_{a}=t^{b}_{\leftarrow}=\frac{1}{2}r_{+}\ ,\nonumber\\ 
&t_{a}=\frac{1}{2}\left(t_{+}+1\right)\ , \nonumber\\
&t^{b}_{\rightarrow}=\frac{1}{2}\left(t_{+}-1\right)\ .
\label{transm}
\end{align}
Reflection, transmission and transfer probabilities are computed as  $R_{a}=\vert r_{a}\vert^{2}$, $T_{a}=\vert t_{a}\vert^{2}$, $T^{b}_{\leftarrow}=\vert t^{b}_{\leftarrow}\vert^{2}$ and $T^{b}_{\rightarrow}=\vert t^{b}_{\rightarrow}\vert^{2}$. The scattering amplitudes satisfy the standard flow conservation condition: $R_{a}+T_{a}+T^{b}_{\leftarrow}+T^{b}_{\rightarrow}=1$.

\begin{figure}[ht]
\includegraphics[width=\columnwidth]{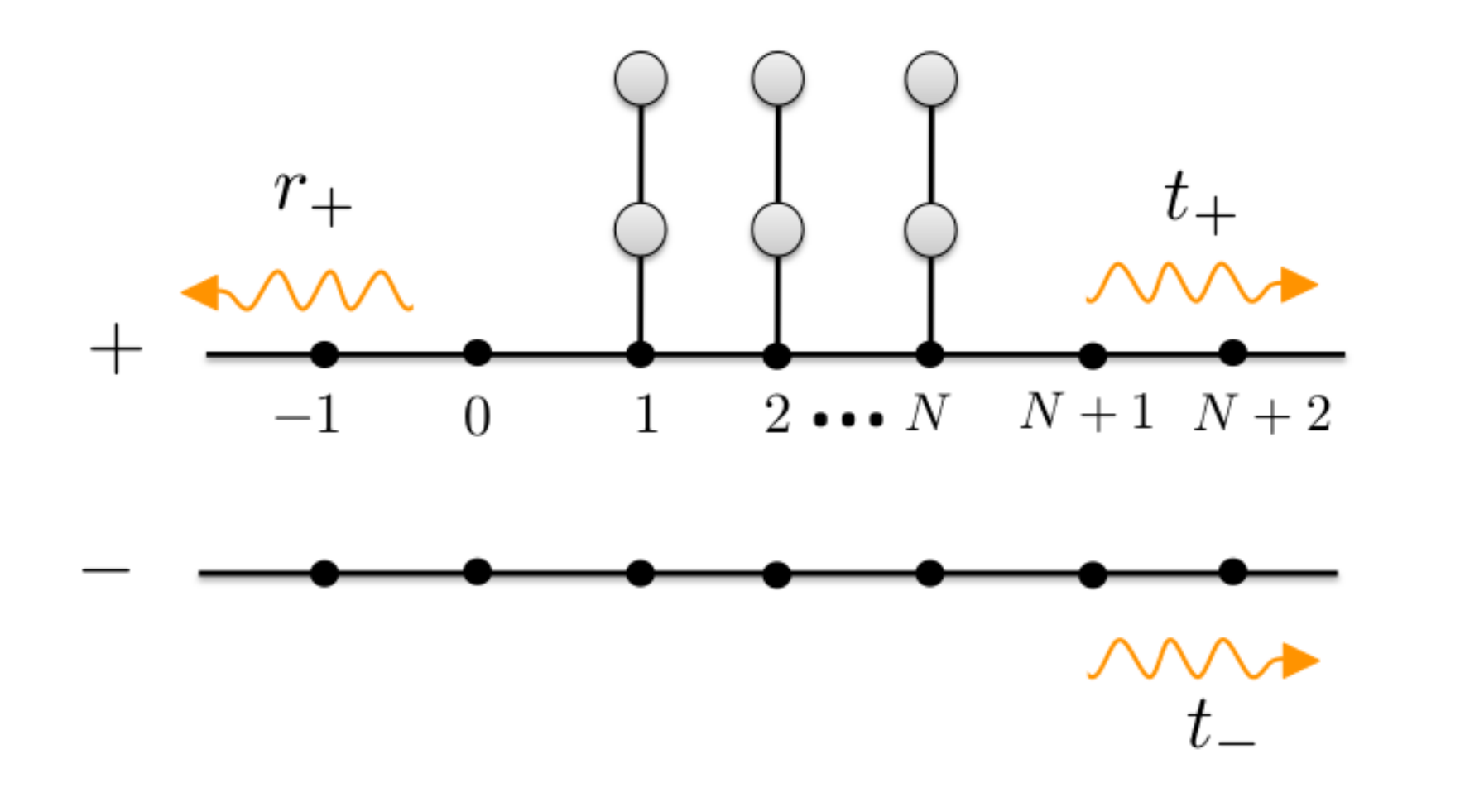}
\caption{Schematic representation of the virtual symmetric and antisymmetric channels labeled with $+$ and $-$, respectively. The amplitudes of the transmission $t_+$, $t_-$ and, reflection $r_+$, are represented by yellow wavy arrows.}
\label{FigGrafEq}
\end{figure}

The model in the $\mathcal{S-A}$ representation is shown schematically in Fig.~\ref{FigGrafEq}. The $\mathcal{S}$-channel is analogous to an array of $N$ nanowires with one or two sites side-coupled to a quantum wire~\cite{ORELLANA2005384} or a CRW with embedded three-level atoms~\cite{AHUMADA2014366}. Note that $\mathcal{S}$ and $\mathcal{A}$ channels are decoupled. The effective site energy $\varepsilon_j^{+}$ of the symmetric channel is renormalized with respect to $\omega_0$ within the scattering region, which results in scattering in such channel. Contrary to that, the site energy remains constant in the $\mathcal{A}$-channel, resulting in the free wave propagation in it. These considerations are crucial for the explanation of some effects that we discuss in the following sections.

\section{Single photon splitting}

First, we address the simplest case of zero control field ($\Omega=0$) when the third states $\vert s_{j}\rangle$ are decoupled from the rest of the system. In this case, an incident photon is scattered by a set of cavities coupled by $N$ \emph{two-level} atoms. 

\begin{figure}[t!]
\includegraphics[width=\columnwidth]
{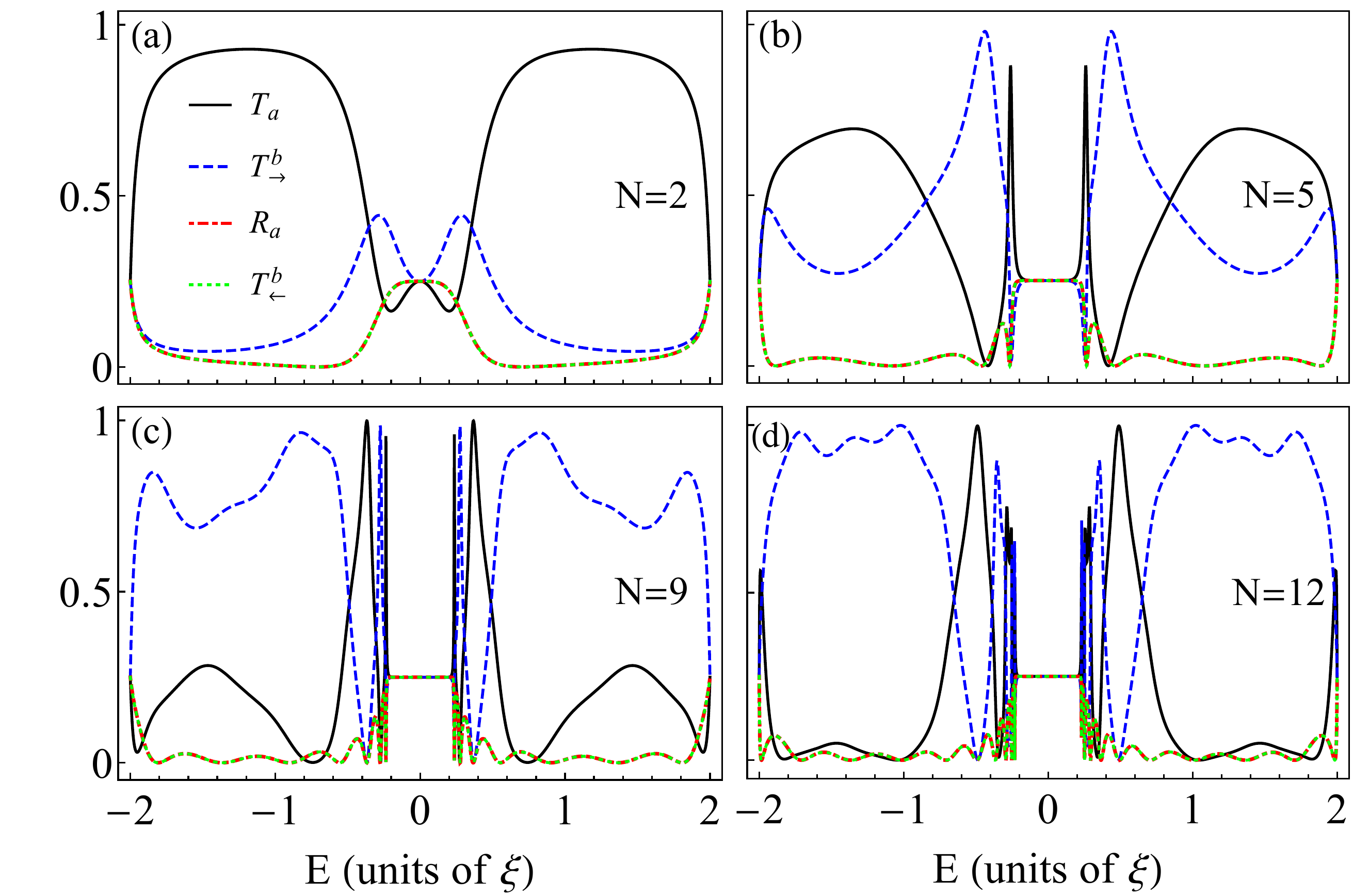}
\caption{Spectra of single photon transmission $T_{a}$ (solid black line), reflection $R_{a}$ (red dashed line), and transfers $T^{b}_{\rightarrow}$ (blue dashed line) and $T^{b}_{\leftarrow}$ (green dotted line). The spectra are calculated for $\Omega=0$, $\omega_{s}=\omega_{e}=\omega_{0}=0$, $g=0.5$ and different numbers of atoms $N$ (specified in the panels).} 
\label{Fig2}
\end{figure}

Figure~\ref{Fig2} shows the transmission, reflection, and transfer probabilities as functions of the incident energy $E$ for the resonant case $\omega_e=\omega_0=0$, $g=0.5$, and different values of the number of atoms $N$. The spectra manifest a very interesting feature: they are degenerate at the center of the band ($E=0$), that is, all four probabilities are equal to $1/4$. The latter equality means that after scattering a photon can leave the system through either of the four channel branches with equal probability. The system can be operating therefore as a perfect ``splitter" of a  photon with this energy. As the number of atoms increases, a flat sub-band is formed about the degeneracy point ($E=\omega_{e}=0$). Within this sub-band the transmissions, reflections, and transfers remain very close to $1/4$. The sub-band is well defined for arrays with $N\geq 3$, its width is growing as $N$ increases and almost saturates for $N=5$.  

Another interesting feature of the transmission spectra in Fig.~\ref{Fig2} is the formation of side-bands of high forward transfer probability into the channel $b$ ($T^{b}_{\rightarrow}$) and, consequently, low transmission probability $T_a$. Figure \ref{Fig2}(b) shows perfect forward transfer into channel $b$ ($T^{b}_{\rightarrow} = 1$) at certain values of the energy of the incident photon: see the transfer peak at $E\approx \pm 0.6$. With increasing values of $N$, two forward transfer sub-bands are formed, as can be seen in Figs.~\ref{Fig2}(c) and~\ref{Fig2}(d) for $E\lesssim -1$ and $E\gtrsim 1$. As can also be seen from Fig.~\ref{Fig2}(d), broad peaks of high transmission $T_a$ (and low transfer $T^{b}_{\rightarrow}$) appear in the vicinity of high transfer sub-bands: see the peaks of $T_a$ at $E\approx \pm 0.5$. Such high forward transfer sub-bands having neighboring peaks of high $T_a$ are very useful for photon routing or switching, which we discuss in the next section.

\begin{figure}[t!]
\includegraphics[width=0.82\linewidth]{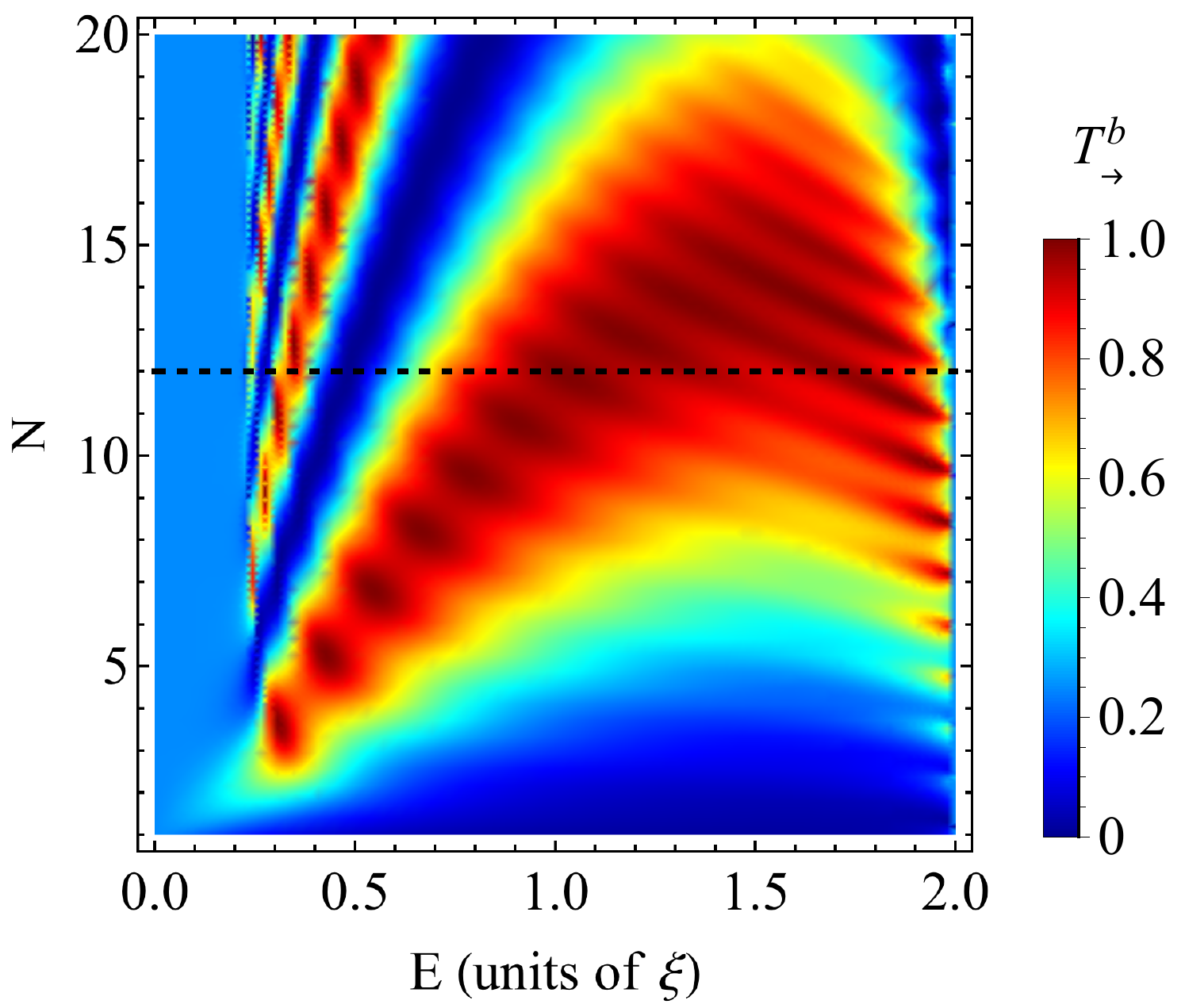}
\caption{Density plot of the probability of forward transfer $T^{b}_{\rightarrow}$ as a function of the incident photon energy $E$ and the number of atoms $N$. The black dashed line indicates the optimal system configuration of $N=12$ atoms. The spectra are calculated for $\omega_{e}=\omega_{s}=\omega_{0}=0$, $g=0.5$, and $\Omega=0$}
\label{FigN}
\end{figure}

Next, we look for a system configuration which would be most appropriate for photon routing. To this end, we show in Figure~\ref{FigN} the transfer spectrum as a function of the photon energy $E>0$ and the number of atoms $N$. Here, the configuration optimal for photon switching seems  to be attained for $N=12$, in which case both the side-band of high transfer (dark red region) and the neighboring sub-band of low transfer (dark blue region) are relatively broad. 

\begin{figure}
\includegraphics[width=\columnwidth]{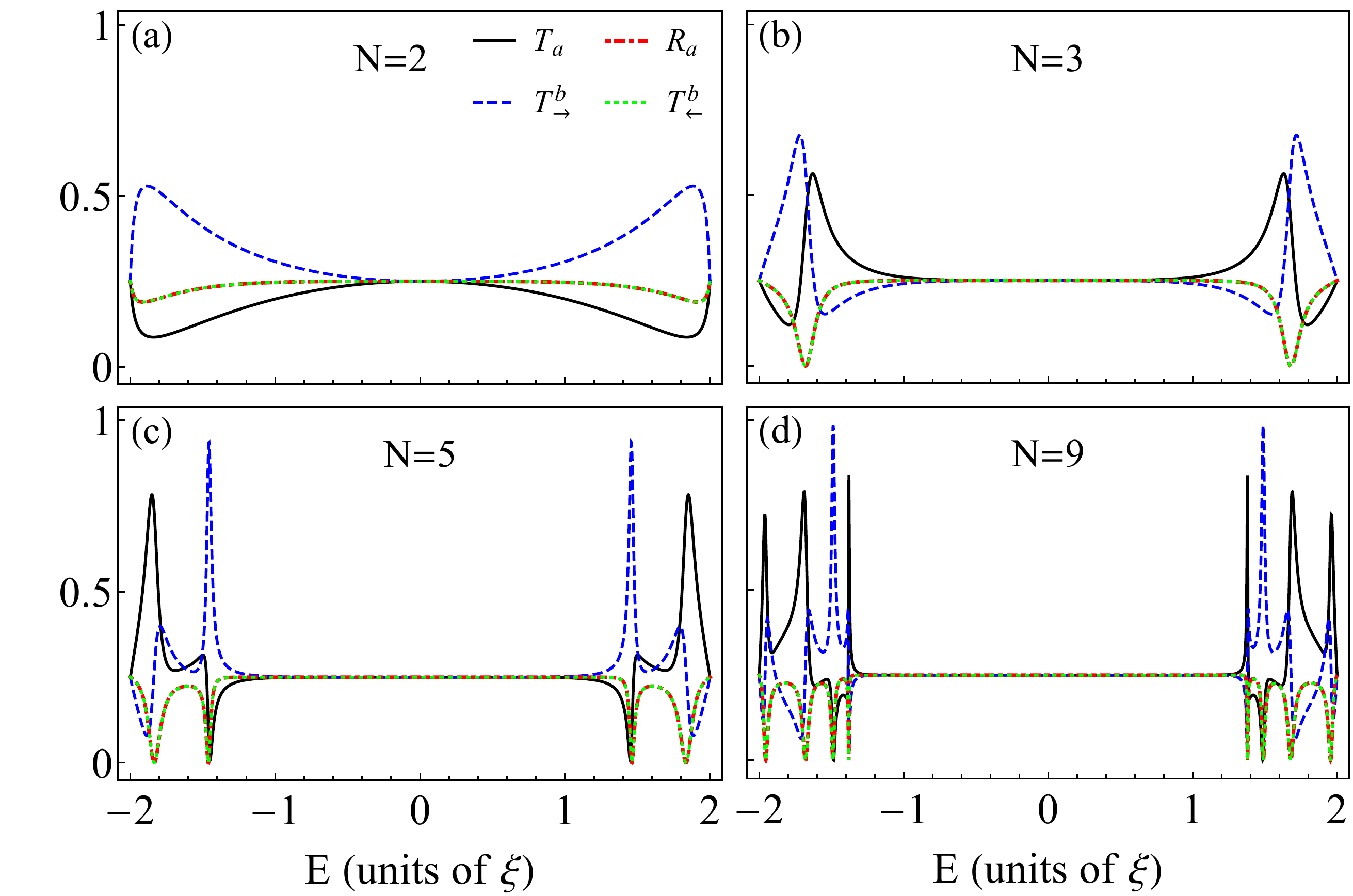}
\caption{Single photon transmission $T_{a}$ (solid black line), reflection $R_{a}$ (red dashed line) and transfer $T^{b}_{\rightarrow}$ (blue dashed line) and  $T^{b}_{\leftarrow}$ (green dotted line) spectra as a function of the incident energy $E$. The spectra are calculated for parameters  $\Omega=0$, $\omega_{0}=\omega_{s}=\omega_{e}=0$,  $g=1.5$, and different numbers of atoms specified in the panels.}
\label{Fig3}
\end{figure}

Figure~\ref{Fig3} shows the transmission, reflection, and transfer spectra as a function of the incident photon energy $E$ for $g=1.5$ and different numbers of the atoms $N$. Here, $T_{a}$, $R_{a}$, $T^{b}_{\leftarrow}$ and, $T^{b}_{\rightarrow}$ remain constant at $1/4$ within a broader sub-band compared to the previous case of $g=0.5$.  As $N$ increases,  the edges of the sub-band become better defined, the bandwidth increases and saturates for $N\geq 5$ at a value on the order of $2g$ [see Fig.~\ref{Fig3} (c) and (d)]. 

\begin{figure}
\includegraphics[width=\columnwidth]{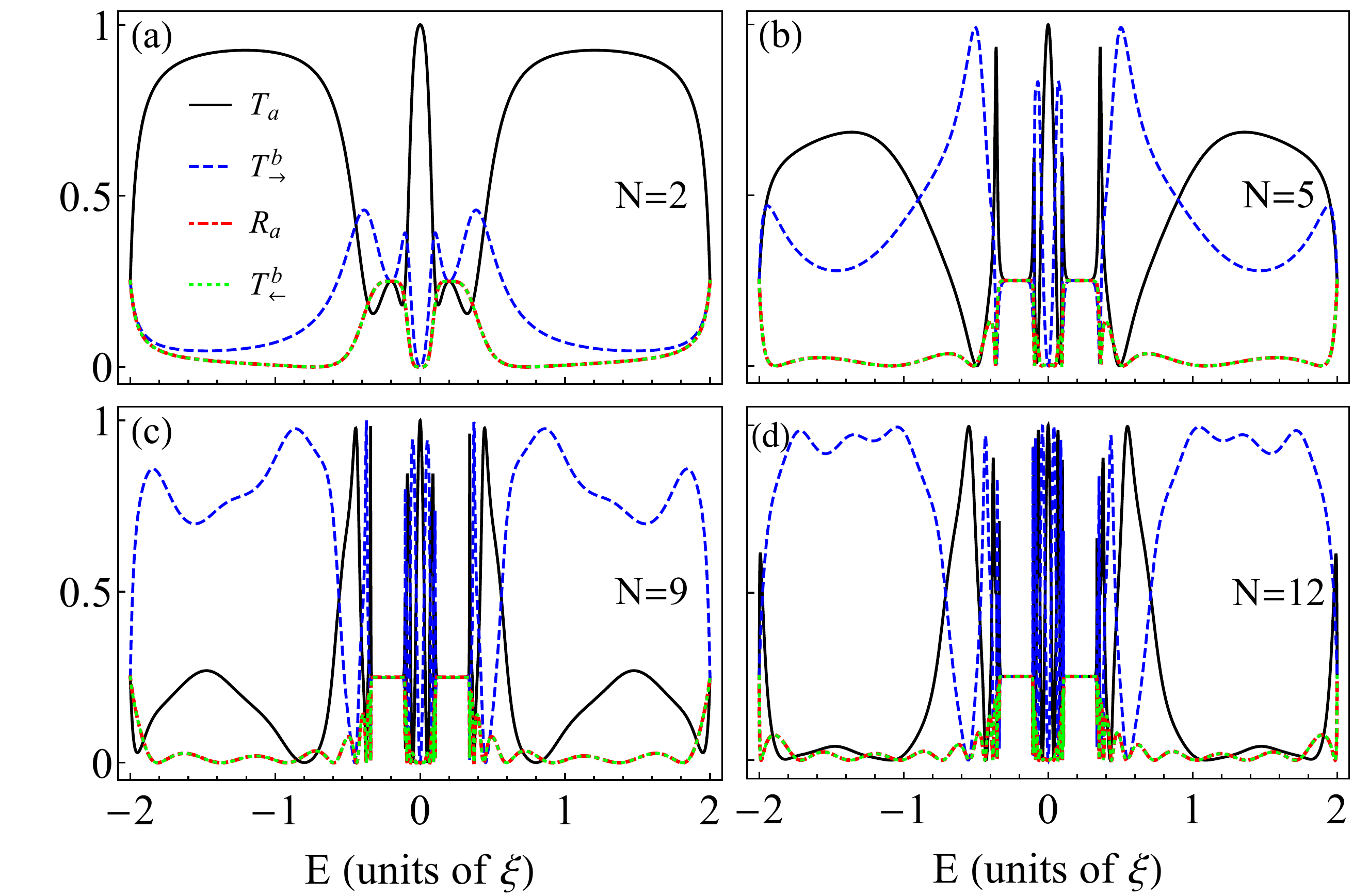}
\caption{Single photon transmission $T_{a}$ (solid black line), reflection $R_{a}$ (red dashed line) and transfer $T^{b}_{\rightarrow}$ (blue dashed line) and  $T^{b}_{\leftarrow}$ (green dotted line) spectra as a function of the incident energy $E$. The spectra are calculated for parameters  $\Omega=0.2$, $\omega_{0}=\omega_{s}=\omega_{e}=0$,  $g=0.5$, and different numbers of atoms specified in the panels.} 
\label{Fig4}
\end{figure}

In the presence of the external electromagnetic field ($\Omega\neq0$), the scattering region comprises sections of the CRWs coupled via $N$ three-level systems. Figure~\ref{Fig4} shows transmission, transfer, and reflection spectra as a function of the photon energy $E$ and different number of atoms $N$ for $\Omega=0.2$. Contrary to the previous case of $\Omega=0$, two flat almost degenerate 1/4 sub-bands are formed. This time they are centered at $E=\pm\Omega$, which suggests that their position can be controlled by the external field. The latter feature is very promising from the point of view of real-time control of photon routing or switching, as we argue below.

Next, we discuss the flat sub-band formation. As we have demonstrated in the previous section, the virtual $\mathcal{S}$ and $\mathcal{A}$ channels are decoupled, as shown schematically in Fig.~\ref{FigGrafEq}. First, the $\mathcal{A}$-channel is equivalent to a free channel with energy $\omega_0$, where the incident wave is transmitted without reflection and with unity transmission amplitude, i.e. $r_{-}=0$ and $t_{-}=1$ then $\vert t_{-}\vert^{2}=T_{-}=1$ and $\vert r_{-}\vert^{2}=R_{-}=0$, which results in the equality $R_{a}=T^{b}_{\leftarrow}$ [see Eqs.~(\ref{transm})].

Second, the effective site energy within the scattering region of the $\mathcal{S}$-channel is given by:
$$
 \varepsilon^{+}_{j}= \varepsilon^{+}(E)=\omega_{0}+2g^2 V(E)\ ,\quad 1 \leq j \leq N\ .
$$
The function $V(E)$ [defined in Eq.~(\ref{V})] has poles $E_\pm$:
\begin{equation}
E_\pm=\frac{\omega _e+\omega _s \pm D}{2}\ ,\quad D=\sqrt{\left(\omega _e-\omega _s\right){}^2+4\, \Omega ^2}\ .
\label{Epm}
\end{equation}
Therefore, the site energy $\varepsilon^{+}(E)$ diverges at the poles and effectively breaks the channel, which results in the total reflection (or zero transmission) in the $\mathcal{S}$- channel. The physical origin of the vanishing transmission is the Fano effect~\cite{PhysRev.124.1866, RevModPhys.82.2257}. A photon has two virtual paths in the $\mathcal{S}$-channel: a direct one without scattering and an indirect path with it. The destructive interference between the two paths results in the zero transmission.\cite{AHUMADA2014366}

In the simplest case $\omega_{0}=\omega_{s}=\omega_{e}=\Omega=0$, the two poles are degenerate and the transmission in the $\mathcal{S}$-channel vanishes at $E=0$. In the most general case, the position of the poles depends on $\omega_e$, $\omega_s$, and $\Omega$.

Finally, if  $|(E-\varepsilon^{+})/2\xi|> 1$, the transmission $t_{+}$ in the $\mathcal{S}$-channel can be rewritten as 
\begin{align}
t_{+}&=\frac{e^{-ikN}\sin k\sinh \kappa_{+}}{D_{+}}\ ,\nonumber \\ 
D_{+}&=\sin k\sinh \kappa_ {+}\cosh(N\kappa_ {+})\nonumber \\
&+i(\cos k\cosh \kappa_{+}-1)\sinh(N\kappa_{+})\ , \nonumber\\
\kappa_{+}&=\cosh^{-1}\left(-\frac{E-\varepsilon^{+}(E)}{2\xi}\right). 
\end{align}
If $E$ is sufficiently close to a pole, $\kappa_+ \gg 1$ and we can approximate $\sinh \kappa_{+}N \simeq \cosh \kappa_{+}N \simeq \exp(\kappa_{+}N)/2$ and consequently the transmission amplitude scales as $$t_{+} \sim \exp(-\kappa_{+}N)\ll 1.$$ As the number of atoms $N$ increases, the region of energies where this approximation is valid becomes broader and the forbidden transmission sub-band is formed in the $\mathcal{S}$-channel. Within this sub-band $t_{+}\rightarrow0$ and $r_{+}\rightarrow1$ and then, from Eqs.~(\ref{transm}), one obtains $r_{a}=t_{a}=t^{b}_{\leftarrow}=-t^{b}_{\rightarrow}\approx 1/2$ or equivalently $R_{a}=T_{a}=T^{b}_{\leftarrow}=T^{b}_{\rightarrow}\approx 1/4$. The latter equality describes the flat bands in Figs.~\ref{Fig2}, \ref{Fig3}, and \ref{Fig4}.

The flat bands are formed about a resonance (or rather anti-resonance) energy $E_\pm$ and their widths are proportional to the atom-to-CRW coupling constant $g$. A physical explanation of the band formation is the following: the strict degeneracy occurs due to the Fano effect only at the resonance whose position is determined, in particular, by the atomic transition energy $\omega_e$ [see Eq.~(\ref{Epm})]. Note that a resonance also exists in the case when only one atom is connecting the two CRWs,~\cite{PhysRevLett.111.103604,PhysRevA.89.013805}. However, as more atoms are added to the system, a band of $N$ almost resonant states is formed about $\omega_e$. This results finally in the formation of broad flat almost degenerate bands. Quite naturally, the atom-to-CRW coupling constant $g$ determines the width of these bands. As the number of atoms $N$ increases, this width grows and saturates at $N\geq 5$. Both effects can be seen in Figures~\ref{Fig2}-\ref{Fig4}. The advantage of using various atoms also becomes clear: instead of a relatively narrow one-atom resonance which can be very sensitive to small variations in parameters or external noise. On the contrary, a system with many atoms provides a broad operational band which can be expected to be more robust to small fluctuations.

\section{Controlled photon routing}
\begin{figure*}
\includegraphics[width=0.75\linewidth]{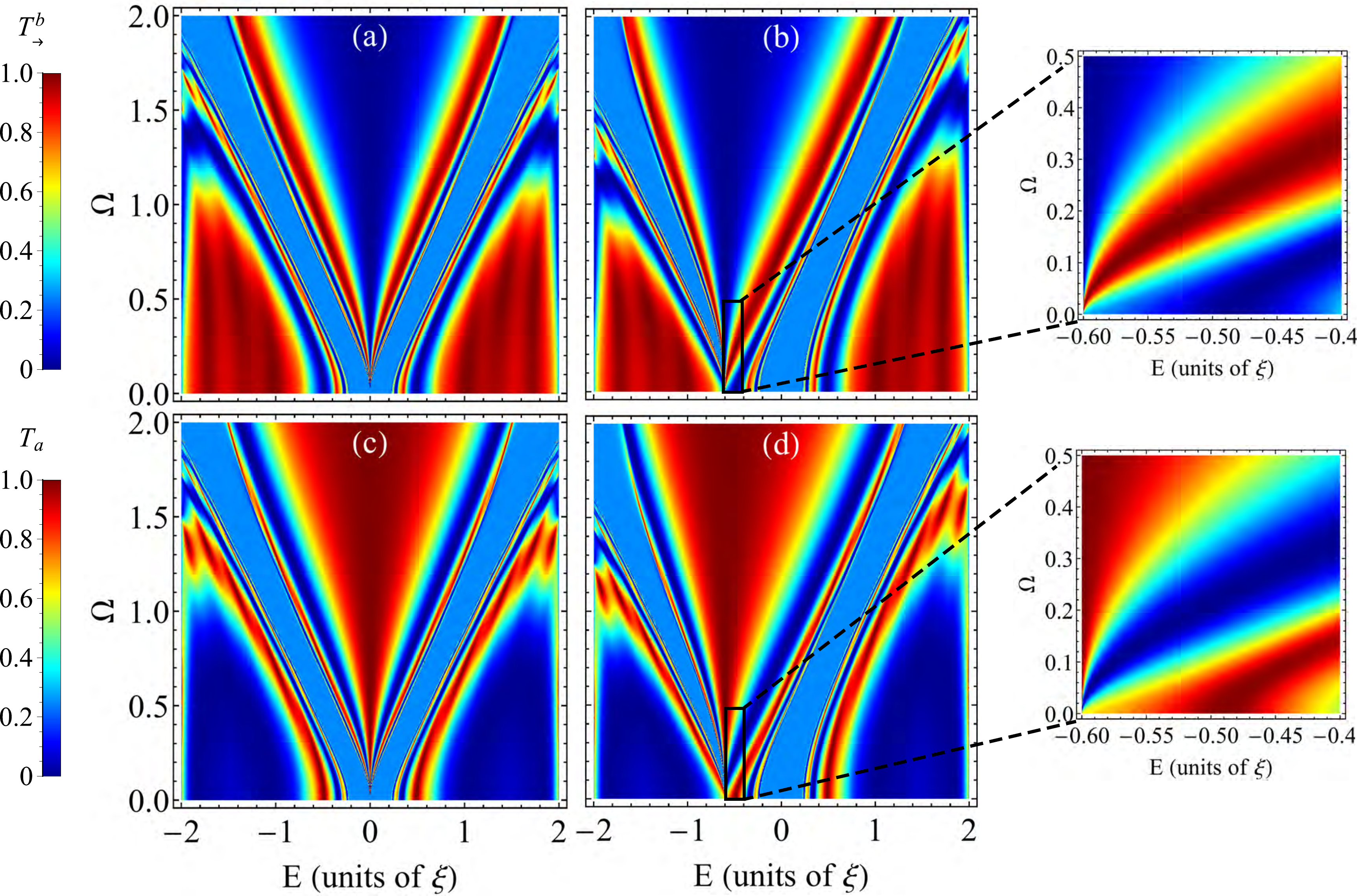}
\caption{Probabilities of transfer $T^{b}_{\rightarrow}$ (upper panels) and transmission $T_{a}$ (lower panels) as a function of photon energy $E$ and the control field Rabi frequency $\Omega$, calculated for $N=12$, $g=0.5$, $\omega_{0}=\omega_{e}=0$, $\omega_{s}=0$ (left column) and $\omega_{s}=-0.6$ (right column).}
\label{Fig5}
\end{figure*}
In this section, we discuss the possibilities of control of the photon propagation. To this end, we show in Figure~\ref{Fig5} the forward transfer probability $T^{b}_{\rightarrow}$ (upper row) and transmission probability $T_a$ (lower row) as a function of photon energy $E$ and the control field $\Omega$ calculated for $N=12$, $\omega_s=0$ (left column), and  $\omega_s=-0.6$ (right column). In Figs.~\ref{Fig5}(a) and \ref{Fig5}(b) we observe two extended transmission sub-bands in $T^{b}_{\rightarrow}$. These regions, indicated in dark red, correspond to a range of values of $E$ and $\Omega$ in which the transfer coefficient $T^{b}_{\rightarrow}$ is unity or nearly unity. A single incident photon with energy within these regions is always transferred to channel $b$. Such range corresponds to blue regions in Figs.~\ref{Fig5}(c) and~\ref{Fig5}(d), where the transmission coefficient $T_{a}$ vanishes, or it is close to zero. This indicates that there is no transmission in the CRW--$a$ within such a region of parameters. In addition, there are two extended regions in Fig.~\ref{Fig5} (indicated in light blue) where the systems act as a single photon splitter.
\begin{figure}
\includegraphics[width=\columnwidth]{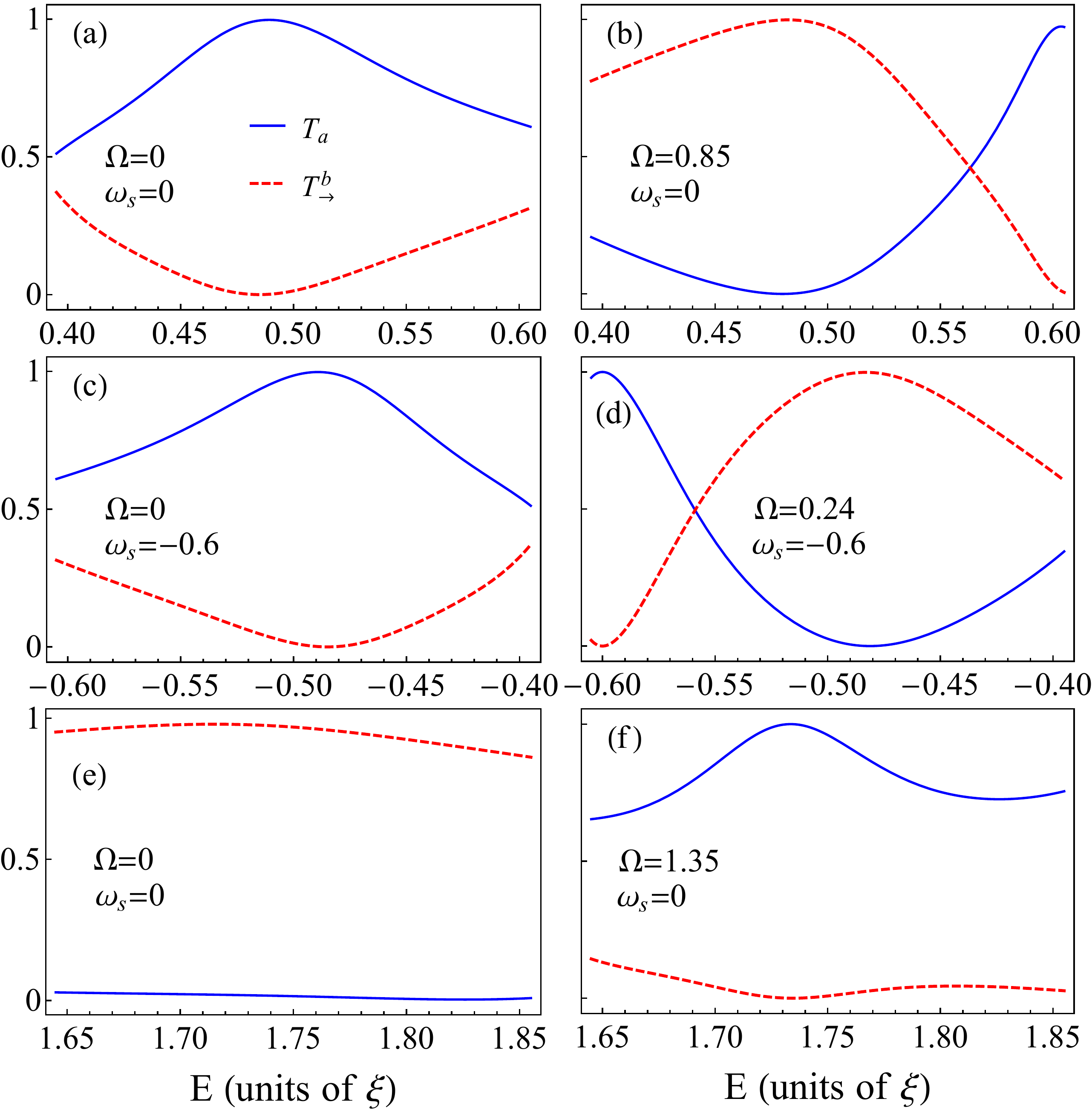}
\caption{Cross sections of Figure~\ref{Fig5}: probabilities of transmission $T_{a}$ (solid blue line) and transfer $T^{b}_{\rightarrow}$ (dashed red line) as a function of the photon energy $E$, calculated for $N=12$, $g=0.5$, $\omega_{0}=\omega_{e}=0$, and different values of the energy of the third atomic level $\omega_{s}$ and control field $\Omega$ (indicated in the panels).}
\label{Fig6}
\end{figure}

Figure~\ref{Fig5} demonstrates the possibility of selecting one of the two physical channels: at some particular energies, the transfer probability $T^{b}_{\rightarrow}$ can be changed from low to high by changing the external field $\Omega$.
That is, when the classical field is turned off ($\Omega=0$), a single photon from channel $a$ exits by this channel with probability unity ($T_{a}=1$) [see the red region in the zoom of Fig.~\ref{Fig5}(d)]. When the external field is applied with characteristics parameters inside the red region of the zoom of Fig.~\ref{Fig5}(b), an incident single-photon can be transferred to the CRW--$b$ with probability $T^{b}_{\rightarrow}=1$. As a consequence, the routing of a single photon from CRW--$a$ to CRW--$b$ can be controlled by the external field.
Control of routing can be seen more clearly in Figure~\ref{Fig6}, where the cross sections of Figure~\ref{Fig5} are shown for the zero (left column) and nonzero (right column) control field $\Omega$. In the upper row of Figure~\ref{Fig6}, the probabilities $T_{a}\approx 1$ and $T^{b}_{\rightarrow}\approx 0$ when the classical field is off ($\Omega=0$), while $T_{a}\approx 0$ and $T^{b}_{\rightarrow}\approx 1$  for the field $\Omega=0.85$. 
Figures~\ref{Fig6}(a) and~\ref{Fig6}(b) show this effect when the energy of the middle state is resonant with the energy of the excited state. This also occurs in Figs.~\ref{Fig6}(c) and~\ref{Fig6}(d), where the energy of the intermediate and excited state are detuned.
A case of the reverse switching is shown in Figs.~\ref{Fig6}(e) and~\ref{Fig6}(f), where $T_{a}\approx 0$ and $T^{b}_{\rightarrow}\approx 1$ when the classical field is turned off, and  $T_{a}\approx 1$ and $T^{b}_{\rightarrow}\approx 0$  when the classical field with $\Omega=1.35$ is on.

The two latter figures suggest that in the degenerate case of $\omega_s=\omega_e=0$, relatively high values of the control field are necessary for routing. However, if the energy of the third atomic level ($\omega_s$) is sufficiently detuned from that of the excited state ($\omega_e$), then the transmission spectra become asymmetric and manifest regions with very inclined alternating bands of high and low transmission for relatively low values of the control field $\Omega$ [see the zooms of Fig.~\ref{Fig5}]. Thus, the transmission spectra can be engineered in such a way that the photon propagation can be controlled by lower classical field, which is generally advantageous. Such a possibility is demonstrated in the middle row of Figure~\ref{Fig6}.

Controlled photon routing or selection of the output channel is possible because all spectral features shift together with the flat bands [see Fig.~\ref{Fig5}]. As we have argued above, positions of the flat bands are determined by the poles $E_\pm$ of the function $V(E)$,  which depend on the Rabi frequency $\Omega$ [see Eq.~(\ref{Epm})]. Thus, by changing the Rabi frequency $\Omega$, the whole spectra can be shifted, switching the system from high transmission to high transfer state or vice versa, controlling the photon propagation.

\section{Conclusions} 

We studied single-photon transport in a system comprising two cavity resonator waveguides coupled via $N$ three-level atoms. One of the allowed atomic transitions is dipole-coupled to resonator modes, while the other -- by an external classical control field. We calculated the transmission, reflection, and transfer spectra for the case of the maximum overlap between the two propagation bands of the waveguides. We showed that the spectra manifest broad flat bands within which an incident photon can scatter and leave the system through either of the four branches of the two channels with equal probability (1/4). Thus, the system can operate as a four-way photon "splitter". The width of the flat bands is determined by the atom-to-waveguide coupling constant, while the positions of the bands depend on the energies of atomic states and the amplitude of the control field. The latter opens a possibility to tune the system, changing its transmission and transfer spectra by the external field, which also means that a photon propagating in the input channel can be routed into one or another output channels selected by the control field. Therefore, the system can operate also as a single-photon switch or router. In comparison with earlier designs of photon routers, our proposed systems have significant improvements, such as a higher routing efficiency and a considerably broader operation bandwidth. Such features make the device more robust and less sensitive to small fluctuations or external noise.

\acknowledgments

Work in Madrid has been supported by MINECO (Grant MAT2016-75955). M.~A. acknowledges financial support from PIIC-UTFSM grant, DGIIP UTFSM, and CONICYT Doctorado Nacional through Grant No. 21141185.

\bibliographystyle{apsrev4-1}
\bibliography{references}

\begin{thebibliography}{30}%
\makeatletter
\providecommand \@ifxundefined [1]{%
 \@ifx{#1\undefined}
}%
\providecommand \@ifnum [1]{%
 \ifnum #1\expandafter \@firstoftwo
 \else \expandafter \@secondoftwo
 \fi
}%
\providecommand \@ifx [1]{%
 \ifx #1\expandafter \@firstoftwo
 \else \expandafter \@secondoftwo
 \fi
}%
\providecommand \natexlab [1]{#1}%
\providecommand \enquote  [1]{``#1''}%
\providecommand \bibnamefont  [1]{#1}%
\providecommand \bibfnamefont [1]{#1}%
\providecommand \citenamefont [1]{#1}%
\providecommand \href@noop [0]{\@secondoftwo}%
\providecommand \href [0]{\begingroup \@sanitize@url \@href}%
\providecommand \@href[1]{\@@startlink{#1}\@@href}%
\providecommand \@@href[1]{\endgroup#1\@@endlink}%
\providecommand \@sanitize@url [0]{\catcode `\\12\catcode `\$12\catcode
  `\&12\catcode `\#12\catcode `\^12\catcode `\_12\catcode `\%12\relax}%
\providecommand \@@startlink[1]{}%
\providecommand \@@endlink[0]{}%
\providecommand \url  [0]{\begingroup\@sanitize@url \@url }%
\providecommand \@url [1]{\endgroup\@href {#1}{\urlprefix }}%
\providecommand \urlprefix  [0]{URL }%
\providecommand \Eprint [0]{\href }%
\providecommand \doibase [0]{http://dx.doi.org/}%
\providecommand \selectlanguage [0]{\@gobble}%
\providecommand \bibinfo  [0]{\@secondoftwo}%
\providecommand \bibfield  [0]{\@secondoftwo}%
\providecommand \translation [1]{[#1]}%
\providecommand \BibitemOpen [0]{}%
\providecommand \bibitemStop [0]{}%
\providecommand \bibitemNoStop [0]{.\EOS\space}%
\providecommand \EOS [0]{\spacefactor3000\relax}%
\providecommand \BibitemShut  [1]{\csname bibitem#1\endcsname}%
\let\auto@bib@innerbib\@empty
%</preamble>
\bibitem [{\citenamefont {Cottet}\ \emph {et~al.}(2017)\citenamefont {Cottet},
  \citenamefont {Dartiailh}, \citenamefont {Desjardins}, \citenamefont
  {Cubaynes}, \citenamefont {Contamin}, \citenamefont {Delbecq}, \citenamefont
  {Viennot}, \citenamefont {Bruhat}, \citenamefont {Dou\c{c}ot},\ and\
  \citenamefont {Kontos}}]{QEDRev}%
  \BibitemOpen
  \bibfield  {author} {\bibinfo {author} {\bibfnamefont {A.}~\bibnamefont
  {Cottet}}, \bibinfo {author} {\bibfnamefont {M.~C.}\ \bibnamefont
  {Dartiailh}}, \bibinfo {author} {\bibfnamefont {M.~M.}\ \bibnamefont
  {Desjardins}}, \bibinfo {author} {\bibfnamefont {T.}~\bibnamefont
  {Cubaynes}}, \bibinfo {author} {\bibfnamefont {L.~C.}\ \bibnamefont
  {Contamin}}, \bibinfo {author} {\bibfnamefont {M.}~\bibnamefont {Delbecq}},
  \bibinfo {author} {\bibfnamefont {J.~J.}\ \bibnamefont {Viennot}}, \bibinfo
  {author} {\bibfnamefont {L.~E.}\ \bibnamefont {Bruhat}}, \bibinfo {author}
  {\bibfnamefont {B.}~\bibnamefont {Dou\c{c}ot}}, \ and\ \bibinfo {author}
  {\bibfnamefont {T.}~\bibnamefont {Kontos}},\ }\href
  {http://stacks.iop.org/0953-8984/29/i=43/a=433002} {\bibfield  {journal}
  {\bibinfo  {journal} {J. Phys. Condens. Matter}\ }\textbf {\bibinfo {volume}
  {29}},\ \bibinfo {pages} {433002} (\bibinfo {year} {2017})}\BibitemShut
  {NoStop}%
\bibitem [{\citenamefont {Xiang}\ \emph {et~al.}(2013)\citenamefont {Xiang},
  \citenamefont {Ashhab}, \citenamefont {You},\ and\ \citenamefont
  {Nori}}]{RevModPhys.85.623}%
  \BibitemOpen
  \bibfield  {author} {\bibinfo {author} {\bibfnamefont {Z.-L.}\ \bibnamefont
  {Xiang}}, \bibinfo {author} {\bibfnamefont {S.}~\bibnamefont {Ashhab}},
  \bibinfo {author} {\bibfnamefont {J.~Q.}\ \bibnamefont {You}}, \ and\
  \bibinfo {author} {\bibfnamefont {F.}~\bibnamefont {Nori}},\ }\href {\doibase
  10.1103/RevModPhys.85.623} {\bibfield  {journal} {\bibinfo  {journal} {Rev.
  Mod. Phys.}\ }\textbf {\bibinfo {volume} {85}},\ \bibinfo {pages} {623}
  (\bibinfo {year} {2013})}\BibitemShut {NoStop}%
\bibitem [{\citenamefont {Reiserer}\ and\ \citenamefont
  {Rempe}(2015)}]{RevModPhys.87.1379}%
  \BibitemOpen
  \bibfield  {author} {\bibinfo {author} {\bibfnamefont {A.}~\bibnamefont
  {Reiserer}}\ and\ \bibinfo {author} {\bibfnamefont {G.}~\bibnamefont
  {Rempe}},\ }\href {\doibase 10.1103/RevModPhys.87.1379} {\bibfield  {journal}
  {\bibinfo  {journal} {Rev. Mod. Phys.}\ }\textbf {\bibinfo {volume} {87}},\
  \bibinfo {pages} {1379} (\bibinfo {year} {2015})}\BibitemShut {NoStop}%
\bibitem [{\citenamefont {Roy}\ \emph {et~al.}(2017)\citenamefont {Roy},
  \citenamefont {Wilson},\ and\ \citenamefont
  {Firstenberg}}]{RevModPhys.89.021001}%
  \BibitemOpen
  \bibfield  {author} {\bibinfo {author} {\bibfnamefont {D.}~\bibnamefont
  {Roy}}, \bibinfo {author} {\bibfnamefont {C.~M.}\ \bibnamefont {Wilson}}, \
  and\ \bibinfo {author} {\bibfnamefont {O.}~\bibnamefont {Firstenberg}},\
  }\href {\doibase 10.1103/RevModPhys.89.021001} {\bibfield  {journal}
  {\bibinfo  {journal} {Rev. Mod. Phys.}\ }\textbf {\bibinfo {volume} {89}},\
  \bibinfo {pages} {021001} (\bibinfo {year} {2017})}\BibitemShut {NoStop}%
\bibitem [{\citenamefont {Notomi}\ \emph {et~al.}(2008)\citenamefont {Notomi},
  \citenamefont {Kuramochi},\ and\ \citenamefont {Tanabe}}]{Notomi2008}%
  \BibitemOpen
  \bibfield  {author} {\bibinfo {author} {\bibfnamefont {M.}~\bibnamefont
  {Notomi}}, \bibinfo {author} {\bibfnamefont {E.}~\bibnamefont {Kuramochi}}, \
  and\ \bibinfo {author} {\bibfnamefont {T.}~\bibnamefont {Tanabe}},\ }\href
  {http://dx.doi.org/10.1038/nphoton.2008.226 http://10.0.4.14/nphoton.2008.226
  https://www.nature.com/articles/nphoton.2008.226{\#}supplementary-information}
  {\bibfield  {journal} {\bibinfo  {journal} {Nat. Photon.}\ }\textbf {\bibinfo
  {volume} {2}},\ \bibinfo {pages} {741} (\bibinfo {year} {2008})}\BibitemShut
  {NoStop}%
\bibitem [{\citenamefont {Monroe}(2002)}]{Monroe2002}%
  \BibitemOpen
  \bibfield  {author} {\bibinfo {author} {\bibfnamefont {C.}~\bibnamefont
  {Monroe}},\ }\href {http://dx.doi.org/10.1038/416238a
  http://10.0.4.14/416238a} {\bibfield  {journal} {\bibinfo  {journal}
  {Nature}\ }\textbf {\bibinfo {volume} {416}},\ \bibinfo {pages} {238}
  (\bibinfo {year} {2002})}\BibitemShut {NoStop}%
\bibitem [{\citenamefont {Northup}\ and\ \citenamefont
  {Blatt}(2014)}]{Northup2014}%
  \BibitemOpen
  \bibfield  {author} {\bibinfo {author} {\bibfnamefont {T.~E.}\ \bibnamefont
  {Northup}}\ and\ \bibinfo {author} {\bibfnamefont {R.}~\bibnamefont
  {Blatt}},\ }\href {http://dx.doi.org/10.1038/nphoton.2014.53
  http://10.0.4.14/nphoton.2014.53} {\bibfield  {journal} {\bibinfo  {journal}
  {Nat. Photon.}\ }\textbf {\bibinfo {volume} {8}},\ \bibinfo {pages} {356}
  (\bibinfo {year} {2014})}\BibitemShut {NoStop}%
\bibitem [{\citenamefont {van Loo}\ \emph {et~al.}(2013)\citenamefont {van
  Loo}, \citenamefont {Fedorov}, \citenamefont {Lalumi{\`e}re}, \citenamefont
  {Sanders}, \citenamefont {Blais},\ and\ \citenamefont
  {Wallraff}}]{vanLoo1494}%
  \BibitemOpen
  \bibfield  {author} {\bibinfo {author} {\bibfnamefont {A.~F.}\ \bibnamefont
  {van Loo}}, \bibinfo {author} {\bibfnamefont {A.}~\bibnamefont {Fedorov}},
  \bibinfo {author} {\bibfnamefont {K.}~\bibnamefont {Lalumi{\`e}re}}, \bibinfo
  {author} {\bibfnamefont {B.~C.}\ \bibnamefont {Sanders}}, \bibinfo {author}
  {\bibfnamefont {A.}~\bibnamefont {Blais}}, \ and\ \bibinfo {author}
  {\bibfnamefont {A.}~\bibnamefont {Wallraff}},\ }\href {\doibase
  10.1126/science.1244324} {\bibfield  {journal} {\bibinfo  {journal}
  {Science}\ }\textbf {\bibinfo {volume} {342}},\ \bibinfo {pages} {1494}
  (\bibinfo {year} {2013})}\BibitemShut {NoStop}%
\bibitem [{\citenamefont {Ritter}\ \emph {et~al.}(2012)\citenamefont {Ritter},
  \citenamefont {N{\"{o}}lleke}, \citenamefont {Hahn}, \citenamefont
  {Reiserer}, \citenamefont {Neuzner}, \citenamefont {Uphoff}, \citenamefont
  {M{\"{u}}cke}, \citenamefont {Figueroa}, \citenamefont {Bochmann},\ and\
  \citenamefont {Rempe}}]{Ritter2012}%
  \BibitemOpen
  \bibfield  {author} {\bibinfo {author} {\bibfnamefont {S.}~\bibnamefont
  {Ritter}}, \bibinfo {author} {\bibfnamefont {C.}~\bibnamefont
  {N{\"{o}}lleke}}, \bibinfo {author} {\bibfnamefont {C.}~\bibnamefont {Hahn}},
  \bibinfo {author} {\bibfnamefont {A.}~\bibnamefont {Reiserer}}, \bibinfo
  {author} {\bibfnamefont {A.}~\bibnamefont {Neuzner}}, \bibinfo {author}
  {\bibfnamefont {M.}~\bibnamefont {Uphoff}}, \bibinfo {author} {\bibfnamefont
  {M.}~\bibnamefont {M{\"{u}}cke}}, \bibinfo {author} {\bibfnamefont
  {E.}~\bibnamefont {Figueroa}}, \bibinfo {author} {\bibfnamefont
  {J.}~\bibnamefont {Bochmann}}, \ and\ \bibinfo {author} {\bibfnamefont
  {G.}~\bibnamefont {Rempe}},\ }\href {http://dx.doi.org/10.1038/nature11023
  http://10.0.4.14/nature11023} {\bibfield  {journal} {\bibinfo  {journal}
  {Nature}\ }\textbf {\bibinfo {volume} {484}},\ \bibinfo {pages} {195}
  (\bibinfo {year} {2012})}\BibitemShut {NoStop}%
\bibitem [{\citenamefont {Kimble}(2008)}]{Kimble2008}%
  \BibitemOpen
  \bibfield  {author} {\bibinfo {author} {\bibfnamefont {H.~J.}\ \bibnamefont
  {Kimble}},\ }\href {http://dx.doi.org/10.1038/nature07127
  http://10.0.4.14/nature07127} {\bibfield  {journal} {\bibinfo  {journal}
  {Nature}\ }\textbf {\bibinfo {volume} {453}},\ \bibinfo {pages} {1023}
  (\bibinfo {year} {2008})}\BibitemShut {NoStop}%
\bibitem [{\citenamefont {Zhou}\ \emph {et~al.}(2013)\citenamefont {Zhou},
  \citenamefont {Yang}, \citenamefont {Li},\ and\ \citenamefont
  {Sun}}]{PhysRevLett.111.103604}%
  \BibitemOpen
  \bibfield  {author} {\bibinfo {author} {\bibfnamefont {L.}~\bibnamefont
  {Zhou}}, \bibinfo {author} {\bibfnamefont {L.-P.}\ \bibnamefont {Yang}},
  \bibinfo {author} {\bibfnamefont {Y.}~\bibnamefont {Li}}, \ and\ \bibinfo
  {author} {\bibfnamefont {C.~P.}\ \bibnamefont {Sun}},\ }\href {\doibase
  10.1103/PhysRevLett.111.103604} {\bibfield  {journal} {\bibinfo  {journal}
  {Phys. Rev. Lett.}\ }\textbf {\bibinfo {volume} {111}},\ \bibinfo {pages}
  {103604} (\bibinfo {year} {2013})}\BibitemShut {NoStop}%
\bibitem [{\citenamefont {Lu}\ \emph {et~al.}(2014)\citenamefont {Lu},
  \citenamefont {Zhou}, \citenamefont {Kuang},\ and\ \citenamefont
  {Nori}}]{PhysRevA.89.013805}%
  \BibitemOpen
  \bibfield  {author} {\bibinfo {author} {\bibfnamefont {J.}~\bibnamefont
  {Lu}}, \bibinfo {author} {\bibfnamefont {L.}~\bibnamefont {Zhou}}, \bibinfo
  {author} {\bibfnamefont {L.-M.}\ \bibnamefont {Kuang}}, \ and\ \bibinfo
  {author} {\bibfnamefont {F.}~\bibnamefont {Nori}},\ }\href {\doibase
  10.1103/PhysRevA.89.013805} {\bibfield  {journal} {\bibinfo  {journal} {Phys.
  Rev. A}\ }\textbf {\bibinfo {volume} {89}},\ \bibinfo {pages} {013805}
  (\bibinfo {year} {2014})}\BibitemShut {NoStop}%
\bibitem [{\citenamefont {Huang}\ \emph
  {et~al.}(2018{\natexlab{a}})\citenamefont {Huang}, \citenamefont {Wang},
  \citenamefont {Wang}, \citenamefont {Li},\ and\ \citenamefont
  {Huang}}]{Huang2018}%
  \BibitemOpen
  \bibfield  {author} {\bibinfo {author} {\bibfnamefont {J.-S.}\ \bibnamefont
  {Huang}}, \bibinfo {author} {\bibfnamefont {J.-W.}\ \bibnamefont {Wang}},
  \bibinfo {author} {\bibfnamefont {Y.}~\bibnamefont {Wang}}, \bibinfo {author}
  {\bibfnamefont {Y.-L.}\ \bibnamefont {Li}}, \ and\ \bibinfo {author}
  {\bibfnamefont {Y.-W.}\ \bibnamefont {Huang}},\ }\href {\doibase
  10.1007/s11128-018-1850-9} {\bibfield  {journal} {\bibinfo  {journal}
  {Quantum Inf. Process.}\ }\textbf {\bibinfo {volume} {17}},\ \bibinfo {pages}
  {78} (\bibinfo {year} {2018}{\natexlab{a}})}\BibitemShut {NoStop}%
\bibitem [{\citenamefont {Huang}\ \emph
  {et~al.}(2018{\natexlab{b}})\citenamefont {Huang}, \citenamefont {Wang},
  \citenamefont {Wang},\ and\ \citenamefont {Zhong}}]{J.Huang}%
  \BibitemOpen
  \bibfield  {author} {\bibinfo {author} {\bibfnamefont {J.-S.}\ \bibnamefont
  {Huang}}, \bibinfo {author} {\bibfnamefont {J.-W.}\ \bibnamefont {Wang}},
  \bibinfo {author} {\bibfnamefont {Y.}~\bibnamefont {Wang}}, \ and\ \bibinfo
  {author} {\bibfnamefont {Y.-W.}\ \bibnamefont {Zhong}},\ }\href
  {http://stacks.iop.org/0953-4075/51/i=2/a=025502} {\bibfield  {journal}
  {\bibinfo  {journal} {J. Phys. B: At., Mol. Opt. Phys.}\ }\textbf {\bibinfo
  {volume} {51}},\ \bibinfo {pages} {025502} (\bibinfo {year}
  {2018}{\natexlab{b}})}\BibitemShut {NoStop}%
\bibitem [{\citenamefont {Lu}\ \emph {et~al.}(2015)\citenamefont {Lu},
  \citenamefont {Wang},\ and\ \citenamefont {Zhou}}]{Lu:15}%
  \BibitemOpen
  \bibfield  {author} {\bibinfo {author} {\bibfnamefont {J.}~\bibnamefont
  {Lu}}, \bibinfo {author} {\bibfnamefont {Z.~H.}\ \bibnamefont {Wang}}, \ and\
  \bibinfo {author} {\bibfnamefont {L.}~\bibnamefont {Zhou}},\ }\href {\doibase
  10.1364/OE.23.022955} {\bibfield  {journal} {\bibinfo  {journal} {Opt.
  Express}\ }\textbf {\bibinfo {volume} {23}},\ \bibinfo {pages} {22955}
  (\bibinfo {year} {2015})}\BibitemShut {NoStop}%
\bibitem [{\citenamefont {Liu}\ and\ \citenamefont {Lu}(2016)}]{Liu2016}%
  \BibitemOpen
  \bibfield  {author} {\bibinfo {author} {\bibfnamefont {L.}~\bibnamefont
  {Liu}}\ and\ \bibinfo {author} {\bibfnamefont {J.}~\bibnamefont {Lu}},\
  }\href {\doibase 10.1007/s11128-016-1479-5} {\bibfield  {journal} {\bibinfo
  {journal} {Quantum Inf. Process.}\ }\textbf {\bibinfo {volume} {16}},\
  \bibinfo {pages} {29} (\bibinfo {year} {2016})}\BibitemShut {NoStop}%
\bibitem [{\citenamefont {Aoki}\ \emph {et~al.}(2009)\citenamefont {Aoki},
  \citenamefont {Parkins}, \citenamefont {Alton}, \citenamefont {Regal},
  \citenamefont {Dayan}, \citenamefont {Ostby}, \citenamefont {Vahala},\ and\
  \citenamefont {Kimble}}]{Aoki2009}%
  \BibitemOpen
  \bibfield  {author} {\bibinfo {author} {\bibfnamefont {T.}~\bibnamefont
  {Aoki}}, \bibinfo {author} {\bibfnamefont {A.~S.}\ \bibnamefont {Parkins}},
  \bibinfo {author} {\bibfnamefont {D.~J.}\ \bibnamefont {Alton}}, \bibinfo
  {author} {\bibfnamefont {C.~A.}\ \bibnamefont {Regal}}, \bibinfo {author}
  {\bibfnamefont {B.}~\bibnamefont {Dayan}}, \bibinfo {author} {\bibfnamefont
  {E.}~\bibnamefont {Ostby}}, \bibinfo {author} {\bibfnamefont {K.~J.}\
  \bibnamefont {Vahala}}, \ and\ \bibinfo {author} {\bibfnamefont {H.~J.}\
  \bibnamefont {Kimble}},\ }\href {\doibase 10.1103/PhysRevLett.102.083601}
  {\bibfield  {journal} {\bibinfo  {journal} {Phys. Rev. Lett.}\ }\textbf
  {\bibinfo {volume} {102}},\ \bibinfo {pages} {083601} (\bibinfo {year}
  {2009})}\BibitemShut {NoStop}%
\bibitem [{\citenamefont {Xia}\ and\ \citenamefont
  {Twamley}(2013)}]{PhysRevX.3.031013}%
  \BibitemOpen
  \bibfield  {author} {\bibinfo {author} {\bibfnamefont {K.}~\bibnamefont
  {Xia}}\ and\ \bibinfo {author} {\bibfnamefont {J.}~\bibnamefont {Twamley}},\
  }\href {\doibase 10.1103/PhysRevX.3.031013} {\bibfield  {journal} {\bibinfo
  {journal} {Phys. Rev. X}\ }\textbf {\bibinfo {volume} {3}},\ \bibinfo {pages}
  {031013} (\bibinfo {year} {2013})}\BibitemShut {NoStop}%
\bibitem [{\citenamefont {Shomroni}\ \emph {et~al.}(2014)\citenamefont
  {Shomroni}, \citenamefont {Rosenblum}, \citenamefont {Lovsky}, \citenamefont
  {Bechler}, \citenamefont {Guendelman},\ and\ \citenamefont
  {Dayan}}]{Shomroni903}%
  \BibitemOpen
  \bibfield  {author} {\bibinfo {author} {\bibfnamefont {I.}~\bibnamefont
  {Shomroni}}, \bibinfo {author} {\bibfnamefont {S.}~\bibnamefont {Rosenblum}},
  \bibinfo {author} {\bibfnamefont {Y.}~\bibnamefont {Lovsky}}, \bibinfo
  {author} {\bibfnamefont {O.}~\bibnamefont {Bechler}}, \bibinfo {author}
  {\bibfnamefont {G.}~\bibnamefont {Guendelman}}, \ and\ \bibinfo {author}
  {\bibfnamefont {B.}~\bibnamefont {Dayan}},\ }\href {\doibase
  10.1126/science.1254699} {\bibfield  {journal} {\bibinfo  {journal}
  {Science}\ }\textbf {\bibinfo {volume} {345}},\ \bibinfo {pages} {903}
  (\bibinfo {year} {2014})}\BibitemShut {NoStop}%
\bibitem [{\citenamefont {Li}\ \emph {et~al.}(2016)\citenamefont {Li},
  \citenamefont {Zhang}, \citenamefont {Xiong},\ and\ \citenamefont
  {Zhou}}]{Li2016}%
  \BibitemOpen
  \bibfield  {author} {\bibinfo {author} {\bibfnamefont {X.}~\bibnamefont
  {Li}}, \bibinfo {author} {\bibfnamefont {W.-Z.}\ \bibnamefont {Zhang}},
  \bibinfo {author} {\bibfnamefont {B.}~\bibnamefont {Xiong}}, \ and\ \bibinfo
  {author} {\bibfnamefont {L.}~\bibnamefont {Zhou}},\ }\href
  {http://dx.doi.org/10.1038/srep39343 http://10.0.4.14/srep39343} {\bibfield
  {journal} {\bibinfo  {journal} {Sci. Rep.}\ }\textbf {\bibinfo {volume}
  {6}},\ \bibinfo {pages} {39343} (\bibinfo {year} {2016})}\BibitemShut
  {NoStop}%
\bibitem [{\citenamefont {Cao}\ \emph {et~al.}(2017)\citenamefont {Cao},
  \citenamefont {Duan}, \citenamefont {Chen}, \citenamefont {Zhang},
  \citenamefont {Wang},\ and\ \citenamefont {Wang}}]{Cao:17}%
  \BibitemOpen
  \bibfield  {author} {\bibinfo {author} {\bibfnamefont {C.}~\bibnamefont
  {Cao}}, \bibinfo {author} {\bibfnamefont {Y.-W.}\ \bibnamefont {Duan}},
  \bibinfo {author} {\bibfnamefont {X.}~\bibnamefont {Chen}}, \bibinfo {author}
  {\bibfnamefont {R.}~\bibnamefont {Zhang}}, \bibinfo {author} {\bibfnamefont
  {T.-J.}\ \bibnamefont {Wang}}, \ and\ \bibinfo {author} {\bibfnamefont
  {C.}~\bibnamefont {Wang}},\ }\href {\doibase 10.1364/OE.25.016931} {\bibfield
   {journal} {\bibinfo  {journal} {Opt. Express}\ }\textbf {\bibinfo {volume}
  {25}},\ \bibinfo {pages} {16931} (\bibinfo {year} {2017})}\BibitemShut
  {NoStop}%
\bibitem [{\citenamefont {Yan}\ and\ \citenamefont {Fan}(2014)}]{Yan2014}%
  \BibitemOpen
  \bibfield  {author} {\bibinfo {author} {\bibfnamefont {W.-B.}\ \bibnamefont
  {Yan}}\ and\ \bibinfo {author} {\bibfnamefont {H.}~\bibnamefont {Fan}},\
  }\href {\doibase 10.1038/srep04820} {\bibfield  {journal} {\bibinfo
  {journal} {Sci. Rep.}\ }\textbf {\bibinfo {volume} {4}},\ \bibinfo {pages}
  {4820} (\bibinfo {year} {2014})}\BibitemShut {NoStop}%
\bibitem [{\citenamefont {Yan}\ \emph {et~al.}(2018)\citenamefont {Yan},
  \citenamefont {Li}, \citenamefont {Yuan},\ and\ \citenamefont
  {Wei}}]{PhysRevA.97.023821}%
  \BibitemOpen
  \bibfield  {author} {\bibinfo {author} {\bibfnamefont {C.-H.}\ \bibnamefont
  {Yan}}, \bibinfo {author} {\bibfnamefont {Y.}~\bibnamefont {Li}}, \bibinfo
  {author} {\bibfnamefont {H.}~\bibnamefont {Yuan}}, \ and\ \bibinfo {author}
  {\bibfnamefont {L.~F.}\ \bibnamefont {Wei}},\ }\href {\doibase
  10.1103/PhysRevA.97.023821} {\bibfield  {journal} {\bibinfo  {journal} {Phys.
  Rev. A}\ }\textbf {\bibinfo {volume} {97}},\ \bibinfo {pages} {023821}
  (\bibinfo {year} {2018})}\BibitemShut {NoStop}%
\bibitem [{\citenamefont {Hoi}\ \emph {et~al.}(2011)\citenamefont {Hoi},
  \citenamefont {Wilson}, \citenamefont {Johansson}, \citenamefont {Palomaki},
  \citenamefont {Peropadre},\ and\ \citenamefont
  {Delsing}}]{PhysRevLett.107.073601}%
  \BibitemOpen
  \bibfield  {author} {\bibinfo {author} {\bibfnamefont {I.-C.}\ \bibnamefont
  {Hoi}}, \bibinfo {author} {\bibfnamefont {C.~M.}\ \bibnamefont {Wilson}},
  \bibinfo {author} {\bibfnamefont {G.}~\bibnamefont {Johansson}}, \bibinfo
  {author} {\bibfnamefont {T.}~\bibnamefont {Palomaki}}, \bibinfo {author}
  {\bibfnamefont {B.}~\bibnamefont {Peropadre}}, \ and\ \bibinfo {author}
  {\bibfnamefont {P.}~\bibnamefont {Delsing}},\ }\href {\doibase
  10.1103/PhysRevLett.107.073601} {\bibfield  {journal} {\bibinfo  {journal}
  {Phys. Rev. Lett.}\ }\textbf {\bibinfo {volume} {107}},\ \bibinfo {pages}
  {073601} (\bibinfo {year} {2011})}\BibitemShut {NoStop}%
\bibitem [{\citenamefont {Yuan}\ \emph {et~al.}(2015)\citenamefont {Yuan},
  \citenamefont {Ma}, \citenamefont {Hou}, \citenamefont {Chang}, \citenamefont
  {Zu},\ and\ \citenamefont {Duan}}]{Yuan2015}%
  \BibitemOpen
  \bibfield  {author} {\bibinfo {author} {\bibfnamefont {X.~X.}\ \bibnamefont
  {Yuan}}, \bibinfo {author} {\bibfnamefont {J.-J.}\ \bibnamefont {Ma}},
  \bibinfo {author} {\bibfnamefont {P.-Y.}\ \bibnamefont {Hou}}, \bibinfo
  {author} {\bibfnamefont {X.-Y.}\ \bibnamefont {Chang}}, \bibinfo {author}
  {\bibfnamefont {C.}~\bibnamefont {Zu}}, \ and\ \bibinfo {author}
  {\bibfnamefont {L.-M.}\ \bibnamefont {Duan}},\ }\href
  {http://dx.doi.org/10.1038/srep12452 http://10.0.4.14/srep12452} {\bibfield
  {journal} {\bibinfo  {journal} {Sci. Rep.}\ }\textbf {\bibinfo {volume}
  {5}},\ \bibinfo {pages} {12452} (\bibinfo {year} {2015})}\BibitemShut
  {NoStop}%
\bibitem [{\citenamefont {Hu}(2017)}]{Hu2017}%
  \BibitemOpen
  \bibfield  {author} {\bibinfo {author} {\bibfnamefont {C.~Y.}\ \bibnamefont
  {Hu}},\ }\href {http://dx.doi.org/10.1038/srep45582
  http://10.0.4.14/srep45582} {\bibfield  {journal} {\bibinfo  {journal} {Sci.
  Rep.}\ }\textbf {\bibinfo {volume} {7}},\ \bibinfo {pages} {45582} (\bibinfo
  {year} {2017})}\BibitemShut {NoStop}%
\bibitem [{\citenamefont {Orellana}\ \emph {et~al.}(2005)\citenamefont
  {Orellana}, \citenamefont {Ladr\'{o}n~de Guevara},\ and\ \citenamefont
  {Dom\'{\i}nguez-Adame}}]{ORELLANA2005384}%
  \BibitemOpen
  \bibfield  {author} {\bibinfo {author} {\bibfnamefont {P.}~\bibnamefont
  {Orellana}}, \bibinfo {author} {\bibfnamefont {M.~L.}\ \bibnamefont
  {Ladr\'{o}n~de Guevara}}, \ and\ \bibinfo {author} {\bibfnamefont
  {F.}~\bibnamefont {Dom\'{\i}nguez-Adame}},\ }\href {\doibase
  https://doi.org/10.1016/j.physe.2004.06.052} {\bibfield  {journal} {\bibinfo
  {journal} {Physica E}\ }\textbf {\bibinfo {volume} {25}},\ \bibinfo {pages}
  {384 } (\bibinfo {year} {2005})}\BibitemShut {NoStop}%
\bibitem [{\citenamefont {Ahumada}\ \emph {et~al.}(2014)\citenamefont
  {Ahumada}, \citenamefont {Cort\'es}, \citenamefont {de~Guevara},\ and\
  \citenamefont {{Ore\-llana}}}]{AHUMADA2014366}%
  \BibitemOpen
  \bibfield  {author} {\bibinfo {author} {\bibfnamefont {M.}~\bibnamefont
  {Ahumada}}, \bibinfo {author} {\bibfnamefont {N.}~\bibnamefont {Cort\'es}},
  \bibinfo {author} {\bibfnamefont {M.~L.}\ \bibnamefont {de~Guevara}}, \ and\
  \bibinfo {author} {\bibfnamefont {P.}~\bibnamefont {{Ore\-llana}}},\ }\href
  {\doibase https://doi.org/10.1016/j.optcom.2014.06.068} {\bibfield  {journal}
  {\bibinfo  {journal} {Opt. Commun.}\ }\textbf {\bibinfo {volume} {332}},\
  \bibinfo {pages} {366 } (\bibinfo {year} {2014})}\BibitemShut {NoStop}%
\bibitem [{\citenamefont {Fano}(1961)}]{PhysRev.124.1866}%
  \BibitemOpen
  \bibfield  {author} {\bibinfo {author} {\bibfnamefont {U.}~\bibnamefont
  {Fano}},\ }\href {\doibase 10.1103/PhysRev.124.1866} {\bibfield  {journal}
  {\bibinfo  {journal} {Phys. Rev.}\ }\textbf {\bibinfo {volume} {124}},\
  \bibinfo {pages} {1866} (\bibinfo {year} {1961})}\BibitemShut {NoStop}%
\bibitem [{\citenamefont {Miroshnichenko}\ \emph {et~al.}(2010)\citenamefont
  {Miroshnichenko}, \citenamefont {Flach},\ and\ \citenamefont
  {Kivshar}}]{RevModPhys.82.2257}%
  \BibitemOpen
  \bibfield  {author} {\bibinfo {author} {\bibfnamefont {A.~E.}\ \bibnamefont
  {Miroshnichenko}}, \bibinfo {author} {\bibfnamefont {S.}~\bibnamefont
  {Flach}}, \ and\ \bibinfo {author} {\bibfnamefont {Y.~S.}\ \bibnamefont
  {Kivshar}},\ }\href {\doibase 10.1103/RevModPhys.82.2257} {\bibfield
  {journal} {\bibinfo  {journal} {Rev. Mod. Phys.}\ }\textbf {\bibinfo {volume}
  {82}},\ \bibinfo {pages} {2257} (\bibinfo {year} {2010})}\BibitemShut
  {NoStop}%
\end{thebibliography}%



\end{document}